\documentclass[fp]{jpsj3}
\usepackage{txfonts}
\usepackage{bm}
\usepackage{graphicx}

\begin{document}

\title{Strong-coupling approach to antiferromagnetic ordering driven by paramagnetic pair-breaking in $d$-wave superconducting phase}

\author{\name{Yuhki \surname{Hatakeyama}}, and \name{Ryusuke \surname{Ikeda}}}
\inst{Department of Physics, Kyoto University, \address{Kyoto 606-8502, Japan}}

\abst{
The field-induced antiferromagnetic (AFM) ordering in the $d_{x^2-y^2}$-paired superconducting phase, which has been recently found in the weak-coupling approach as a basic mechanism due to the Pauli paramagnetic pair-breaking (PPB) in relation to the high-field behaviors in CeCoIn$_5$, is studied in the strong-coupling approach taking account of the electron correlation. Applying the fluctuation-exchange (FLEX) approximation to the two-dimensional Hubbard model including the Zeeman term, it is shown that the PPB-induced AFM ordering in the superconducting (SC) phase and the first order SC transition on $H_{c2}(T)$ are realized in the strong-coupling approach as well as those in the weak-coupling model, and that the AFM ordering is affected by the quasiparticle renormalization and the amplitude of the SC order parameter. This AFM ordering may appear in a wide range of materials close to an AFM quantum critical point (QCP). 
}

\kword{antiferromagnetism, strong-coupling superconductivity, paramagnetic pair-breaking, two-dimensional Hubbard model, FLEX approximation}

\maketitle

\section{
Introduction
\label{sec:introduction}
}
Recently, a field-induced antiferromagnetism inside a superconducting (SC) phase has been observed or suggested 
in many $d_{x^2-y^2}$-wave superconductors with strong paramagnetic pair-breaking (PPB) effect. 
In a heavy electron superconductor CeCoIn$_5$, 
a novel high-field low-temperature (HFLT) SC state has been detected in the high-field region of the SC phase in $\bm{H} \perp \text{c}$, i.e., in the field parallel to the basal plane \cite{bianchi_possible_2003}. Strong indications of the long-sought Fulde-Ferrell-Larkin-Ovchinnikov (FFLO) SC state \cite{fulde_superconductivity_1964_larkin_inhomogeneous_1965} in the HFLT phase have been found through the anomalously strong impurity effect\cite{tokiwa_anisotropic_2008,tokiwa_anomalous_2010} and the observation of normal quasiparticles by NMR experiment\cite{kumagai_evolution_2011}. 
On the other hand, the neutron scattering experiment\cite{kenzelmann_coupled_2008_kenzelmann_evidence_2010} and NMR measurement 
\cite{kumagai_evolution_2011} show that the 
AFM order is present only in the HFLT phase and \textit{not} present in the high-field \textit{normal} phase. This antiferromagnetic (AFM) ordering is quite unusual because the conventional wisdom is that an AFM ordering is competitive with a SC order and tends to be suppressed in the SC phase. 
A field-induced enhancement of an AFM order in the $d$-wave SC phase is also suggested in pressurized CeRhIn$_5$\cite{park_hidden_2006}. 
In addition, the existence of an AFM quantum critical point (QCP) located slightly below $H_{c2}(T=0)$ 
is suggested in many superconductors with strong PPB effect 
such as CeCoIn$_5$ in ${\bf H} \parallel c$ \cite{bianchi_avoided_2003,paglione_field-induced_2003,singh_probing_2007,zaum_towards_2011}, 
CeRhIn$_5$\cite{park_hidden_2006,park_field-induced_2010}, 
Ce$_2$PdIn$_8$\cite{dong_field-induced_2011,tokiwa_quantum_2011}, and NpPd$_2$Al$_5$\cite{honda_effect_2008}. 
This indicates that the AFM critical fluctuations in these materials are the strongest in the high-field region {\it of the SC phase}. 
These experimental results suggest that there is a universal mechanism 
to enhance the AFM ordering or fluctuation in a high-field $d$-wave SC phase with a strong PPB effect.

Recently, we have argued\cite{hatakeyama_emergent_2011} that such a field-induced enhancement of antiferromagnetism, seen commonly in many superconductors, can be explained based on the mechanism of the PPB-induced AFM ordering found in the weak-coupling BCS model\cite{ikeda_antiferromagnetic_2010}. In the high-field region of a $d_{x^2-y^2}$-wave SC phase with strong PPB, the AFM order tends to be formed more easily than in the normal state\cite{ikeda_antiferromagnetic_2010}, and the resulting modulation wave vector $\bm{Q}$ of the AFM order is directed to a gap node of the $d_{x^2-y^2}$-paired gap function. This result can also account for the field-induced enhancement of the AFM ordering in CeRhIn$_5$ 
and the enhancement of the AFM critical fluctuations slightly below $H_{c2}(0)$ in the superconductors with strong PPB effect on the same footing. 
Moreover, if a spatial modulation of the SC order parameter of the Larkin-Ovchinnikov (LO) type is present, this AFM order is stabilized further by this modulation of the SC order\cite{hatakeyama_emergent_2011}. In this manner, the HFLT phase of CeCoIn$_5$ is naturally understood as a coexistent phase of the FFLO and AFM orders induced by PPB. 

While the above-mentioned picture is based on the weak-coupling analysis for the BCS-like (mean field) model for the SC and AFM orders in which the effects of the electron correlation are neglected, most of SC states with strong PPB should occur in materials with strong correlation where $k_{\rm B} T_c/E_{\rm F}$ is not small, because the so-called Maki parameter $\alpha_{\rm M}=H_{orb}/\sqrt{2}H_P$ \cite{maki_effect_1966} ($H_{orb}$ is the orbital depairing field, and $H_P$ is the Pauli limiting field) measuring the strength of PPB is of the order of $k_{\rm B} T_c/E_{\rm F}$, which is assumed to be quite small in the weak-coupling BCS theory. Therefore, one needs to extend the theory to the strong-coupling model to describe those SC behaviors in high fields mentioned above. For instance, the specific heat jump $\Delta C$ at the SC phase transition is larger than that of the weak-coupling BCS model $\Delta C/C_N=1.4$ in CeCoIn$_5$\cite{petrovic_heavy-fermion_2001} and NpPd$_2$Al$_2$\cite{aoki_unconventional_2007}, where $C_N$ is the specific heat in the normal state at $T_c$. 
In addition, when studying the situations near an AFM-QCP, it will be necessary to incorporate effects of the AFM {\it fluctuations}, which are ignored in the mean field description and in the weak-coupling approach, on the mechanism of the AFM ordering in the SC phase. 

In the present paper, we study the strong-coupling effect on the PPB-induced AFM ordering in the $d$-wave SC phase by examining the two-dimensional Hubbard model with the Zeeman energy on the basis of the fluctuation-exchange (FLEX) approximation. 
The two-dimensional Hubbard model is employed because it is the simplest model which can include critical AFM fluctuations. 
The PPB effect is incorporated by including the Zeeman energy in the model Hamiltonian. In the FLEX approximation, the AFM fluctuation in the normal state induces the $d_{x^2-y^2}$-wave pairing and consistently becomes the source of the AFM order induced by the resulting $d_{x^2-y^2}$-wave SC order with the Zeeman energy. 

It is found by taking account of the electronic structures through the FLEX approximation that the PPB-induced AFM ordering is also realized in the  present strong-coupling approach according to almost the same mechanism as in the weak-coupling approach. In addition, 
we investigate the influence of the strong-coupling effect on the high-field AFM phase by assuming a situation near an AFM-QCP. We find that, when the AFM-QCP is approached, the PPB-induced AFM ordering is affected through the following three different effects, 
the quasiparticle renormalization, i.e., the mass enhancement, the $d_{x^2-y^2}$-wave pairing interaction due to the AFM fluctuation, 
and the Stoner enhancement in the normal state. For instance, the quasiparticle renormalization tends to suppress the AFM ordering, while the strong pairing interaction enhances it. However, promotion of the PPB-induced AFM ordering due to the Stoner enhancement is found to dominate over the remaining two effects occurring near the 
AFM-QCP. The sum of the three effects on approaching the AFM-QCP is found to result in an enhancement of the PPB-induced AFM ordering. Therefore, it is expected that the AFM order existing only in the high-field region of the SC state is realized close to the AFM-QCP. 

We also study the SC transition on the $H_{c2}(T)$-line between the uniform SC state and the normal state in high fields. As in the weak-coupling case, the SC ordering from the high-field normal phase occurs as a first order transition (FOT) in the high field and low temperature regime in the present strong-coupling model. 
Consequently, with increasing the field, the PPB-induced AFM order discontinuously vanishes above the $H_{c2}$-line, reflecting the sudden vanishing of the SC energy gap. 

This paper is organized as follows.
In Sect. \ref{sec:model}, we will introduce the model and the method of our numerical calculation based on the FLEX approximation. 
In Sect. \ref{sec:first_order}, we will discuss the FOT between the uniform SC state and the normal phase. 
In Sect. \ref{sec:afm}, we will examine the mechanism of the PPB-induced AFM ordering in the strong-coupling model. 
In Sect. \ref{sec:detail}, we will investigate the influence of the strong-coupling effects on the PPB-induced AFM ordering. 
In Sect. \ref{sec:summary}, we will present the discussion and the summary of the conclusions. Throughout the present paper, the unit $\hbar=k_B=1$ will 
be used.

\section{
Model and method
\label{sec:model}
}
We started from the two-dimensional Hubbard model including the Zeeman energy.
The model Hamiltonian $\mathcal{H}$ is written as
\begin{equation}
\label{eq:Hamiltonian}
	\mathcal{H}=-t_1\sum_{\langle i,j \rangle ,\sigma}\left(c_{i,\sigma}^{\dagger}c_{j,\sigma}+\mathrm{H.c.} \right)
	-t_2\sum_{\langle\langle i,j \rangle\rangle ,\sigma}\left(c_{i,\sigma}^{\dagger}c_{j,\sigma}+\mathrm{H.c.} \right)
	-\sum_{i,\sigma} \left( \mu+h\sigma \right) n_{i,\sigma}+U\sum_i n_{i,\uparrow}n_{i,\downarrow} ,
\end{equation}
where $\sigma(=\pm 1)$ denote the spin projection parallel to the magnetic field, 
and $h=g\mu_B H/2$ is the Zeeman energy ($g$ is a g-factor and $\mu_B$ is the Bohr magneton). 
Here and below, we assume that the PPB effect is so strong that the orbital depairing effect can be ignored, i.e., the Maki parameter $\alpha_M \rightarrow \infty$. 
Further, we will use the lattice spacing $a_l$ ($=1$) and the nearest-neighbor hopping $t_1$ ($=1$) as the unit of length and energy, respectively. Therefore, $H$ is the Zeeman energy or the magnetic field normalized by $t_1$, and $T$ is the dimensionless temperature normalized by $t_1$. 

Regarding the dispersion relation $\epsilon_{\bf k}$ of the free electron, we use that of the two-dimensional square lattice with the next-nearest-neighbor hopping $t_2$ to describe the quasi two-dimensional crystal structure of CeCoIn$_5$ and CeRhIn$_5$. Effects of the third dimension perpendicular to the basal plane will be incorporated later. Then, $\epsilon_{\bf k}$ for this lattice 
structure is given by 
\begin{equation}
	\epsilon_{\bm{k}}=-2t_1 \left( \cos(k_x)+\cos(k_y) \right) -4t_2\cos(k_x)\cos(k_y) .
	\label{eq:dispersion}
\end{equation} 
Close to the half-filling, the Fermi surface of this dispersion relation is nested in the diagonal direction ($1$, $1$) with the nesting 
vector $\bm{Q}\sim(\pi,\pi)$. 
This nesting leads to AFM fluctuation and $d_{x^2-y^2}$-wave superconductivity mediated by this AFM fluctuation.

In the theory of the strong-coupling superconductivity, the electronic state is described by the Nambu Green's function $\hat{G}(k)$,
which is written as
\begin{eqnarray}
\label{eq:dyson}
	\hat{G}(k)	&=& \left( \begin{array}{cc} G_{\uparrow}(k) & F(k) \\ F^{\dagger}(k) & -G_{\downarrow}(-k) \end{array} \right) \nonumber\\
									&=&  \left[ \left(\hat{G}^{(0)}(k) \right)^{-1}-\hat{\Sigma}(k) \right]^{-1} ,
\end{eqnarray}
where $k=(i\omega_n,\bm{k})$ ($\omega_n=(2n+1)\pi T$ is the Fermion Matsubara frequency), and $F^{\dagger}(k)=F^{*}(-k)$.
Here, the non-interacting Green's function $\hat{G}^{(0)}(k)$ and the self-energy $\hat{\Sigma}(k)$ are written as 
\begin{eqnarray}
	\hat{G}^{(0)}(k)	&=& {\left( \begin{array}{cc} i\omega_n-\epsilon_{\bm{k}}+\mu+h & 0 \\ 0 & i\omega_n+\epsilon_{\bm{k}}-\mu+h \end{array} \right)}^{-1} \\
	\hat{\Sigma}(k)	&=& \left( \begin{array}{cc} \Sigma_{\uparrow}(k) & -\Delta(k) \\ -\Delta^{\dagger}(k) & -\Sigma_{\downarrow}(-k) \end{array} \right) ,
\end{eqnarray}
where $\Sigma_{\sigma}(k)$ is the normal self-energy, and $\Delta(k)$ ($\Delta^{\dagger}(k)=\Delta^{*}(-k)$) is the anomalous self-energy.

To obtain the self-energy ${\hat \Sigma}(k)$,we have examined the model Hamiltonian $\mathcal{H}$ using the FLEX approximation, in which the interaction mediated by the AFM fluctuation is incorporated. The crucial point is that the AFM fluctuation and the self-energy are {\it self-consistently} 
calculated. 
This FLEX approximation give a qualitatively good description of the AFM critical fluctuation\cite{moriya_spin_2000} 
and the strong-coupling SC where the pairing interaction is mediated by the AFM fluctuation\cite{monthoux_self-consistent_1994}.

The FLEX approximation can be naturally extended to the case where the time-reversal symmetry is broken by the Zeeman field 
\cite{sakurazawa_magnetic-field-induced_2005}.
Since the FLEX approximation is a kind of the conserving approximation,
the self-consistent equations for the self-energy ${\hat \Sigma}(k)$ are derived from the generating functional $\Phi \left[ G_{\sigma},F,F^{\dagger} \right]$\cite{baym_self-consistent_1962}. 
The generating functional of the FLEX approximation for the model Hamiltonian $\mathcal{H}$ is given by 
\begin{eqnarray}
	\Phi \left[ G_{\sigma},F,F^{\dagger} \right] &=& \sum_{q} \left[ \ln\left[1-U\chi_{\pm}^{(0)}(q)\right]
	\right. \nonumber\\ 
	&& \left. +\frac{1}{2}\ln\left[(1-U\chi_F^{(0)}(q))(1-U\chi_F^{(0)}(-q))-U^2\chi_{\uparrow}^{(0)}(q)\chi_{\downarrow}^{(0)}(q)\right]
	\right. \nonumber\\ 
	&& \left. +\frac{U^2}{2}\left(\chi_F^{(0)}(q)\chi_F^{(0)}(-q)+\chi_{\uparrow}^{(0)}(q)\chi_{\downarrow}^{(0)}(q)\right)+U\chi_{\pm}^{(0)}(q) \right] ,
\end{eqnarray}
where the irreducible susceptibilities $\chi_{\sigma}^{(0)}(q)$ ($\sigma=\uparrow$,$\downarrow$), $\chi_{F}^{(0)}(q)$, and $\chi_{\pm}^{(0)}(q)$ are given by
\begin{eqnarray}
\label{eq:chi}
	\chi_{\sigma}^{(0)}(q) &=& -T\sum_k G_{\sigma}(k)G_{\sigma}(k+q) \\
\label{eq:chiF}
	\chi_{F}^{(0)}(q) &=&-T\sum_k F(k)F^{\dagger}(k+q) \\
\label{eq:chiud}
	\chi_{\pm}^{(0)}(q) &=& -T\sum_k \left[ G_{\uparrow}(k)G_{\downarrow}(k+q)+F(k)F^{\dagger}(-k-q) \right] .
\end{eqnarray}
The self-consistent equations for ${\hat \Sigma}(k)$ are obtained from the functional derivative of $\Phi$ as
\begin{eqnarray}
\label{eq:selfeq_S}
	\Sigma_{\sigma}(k) &=& \frac{\delta \Phi}{\delta G_{\sigma}} \nonumber\\
		&=& T\sum_{q} \left[ U^2\chi_{-\sigma}^{(0)}(q)\left( \frac{1}{(1-U\chi_F^{(0)}(q))(1-U\chi_F^{(0)}(-q))-U^2\chi_{\uparrow}^{(0)}(q)\chi_{\downarrow}^{(0)}(q)}-1 \right) G_{\sigma}(k+q)
		\right. \nonumber\\ 
		&& \left. +\frac{U^2\chi_{\pm}^{(0)}(\sigma q)}{1-U\chi_{\pm}^{(0)}(\sigma q)}G_{-\sigma}(k+q) \right]
\end{eqnarray}
\begin{eqnarray}
\label{eq:selfeq_D}
	\Delta(k) &=& -\frac{\delta \Phi}{\delta F^{\dagger}} \nonumber\\
		&=& -T\sum_{q} \left[ \left( \frac{U(1-U\chi_F^{(0)}(q))}{(1-U\chi_F^{(0)}(q))(1-U\chi_F^{(0)}(-q))-U^2\chi_{\uparrow}^{(0)}(q)\chi_{\downarrow}^{(0)}(q)}-U^2\chi_F^{(0)}(q) \right) F(k+q)
		\right. \nonumber\\ 
		&& \left. +\frac{U^2\chi_{\pm}^{(0)}(q)}{1-U\chi_{\pm}^{(0)}(q)}F(-k-q) \right] .
\end{eqnarray}
We note that the other self-consistent equation
\begin{equation}
	\Delta^{\dagger}(k)=-\frac{\delta \Phi}{\delta F} 
\end{equation}
is equivalent to eq. (\ref{eq:selfeq_D}).
In order to obtain the self-energy $\hat{\Sigma}(k)$ numerically, eqs.(\ref{eq:selfeq_S}) and (\ref{eq:selfeq_D}) have to be self-consistently solved, together with eqs. (\ref{eq:dyson}), (\ref{eq:chi}), (\ref{eq:chiF}) 
and (\ref{eq:chiud}) 
(regarding the calculation in the case without SC order, see Appendix). 

We fix the filling factor $n$ in the self-consistent calculation by adjusting the chemical potential $\mu$ so that the equation
\begin{equation}
	0=-T\sum_{k,\sigma} \left[ G_{\sigma}(k;\mu)-G_{\sigma}^{(0)}(k;\mu_0) \right]
\end{equation}
is satisfied.
Here, $\mu_0$ is the chemical potential in the non-interacting system,
obtained from the equation
\begin{equation}
	1-\delta=\sum_{\bm{k},\sigma} f(\epsilon_{\bm{k}}-\mu_0-h\sigma) ,
\end{equation}
where $\delta$ ($=1-n$) denote the hole doping, and $f(\epsilon)$ is the Fermi distribution function.

The FFLO state is expected to be realized in the high-field region of a SC phase in the strong PPB limit\cite{adachi_effects_2003}. Since the FFLO state is destroyed even by quite a weak impurity effect\cite{ikeda_impurity-induced_2010}, however, 
the FFLO state is not necessarily realized in real $d$-wave superconductors with strong PPB effect. 
Moreover, the PPB-induced AFM ordering mechanism is universal in the $d_{x^2-y^2}$-wave SC state and may be realized even with no FFLO state.\cite{ikeda_antiferromagnetic_2010} 
Therefore, in order to investigate the PPB-induced AFM ordering in general situations, we will focus on the uniform SC state throughout the present paper. 

Because of the strong PPB effect, the FOT from a SC state to the normal state is also expected\cite{maki_pauli_1964}. 
The FOT line is determined by comparing the free energies in the SC and normal states with each other. 
The free energy in the conserving approximation is given by\cite{baym_self-consistent_1962,yanase_kinetic_2005} 
\begin{equation}
\label{eq:FE}
	F(T,n)=T\left\{ \Phi-\sum_{k} \left[ \mathrm{tr}[\hat{\Sigma}\hat{G}] -\ln(-\mathrm{det}[\hat{G}]) \right] \right\}+\mu(n) n .
\end{equation}
In writing this expression, we have performed the Legendre transformation between $\mu$ and $n$ so that $F(T,n)$ is stationary when the filling factor is fixed. That is, the solution of the self-consistent equations (\ref{eq:selfeq_S}), (\ref{eq:selfeq_D}) satisfies
\begin{equation}
	\left( \frac{\delta F}{\delta \Sigma_{\sigma}} \right)_{n}=\left( \frac{\delta F}{\delta \Delta}\right )_{n}=0 .
\end{equation}

As is seen in Sect. \ref{sec:first_order}, we have a FOT between the SC and the normal phases at low temperatures. Because the self-consistent equations (\ref{eq:selfeq_S}) and (\ref{eq:selfeq_D}) have multiple solutions close to this FOT line, 
attention has to be paid in performing the numerical calculations so that
approximate solutions to be used in order for the self-consistent iteration not to go out of the convergence region of the desired solution. 
In our numerical calculation, we employed a linear mixing of approximate solutions in the self-consistent iteration for stable convergence:
$\hat{\Sigma}_{n+1}^{\text{new}}=\beta\hat{\Sigma}_{n+1}+(1-\beta)\hat{\Sigma}_n$, 
where $\hat{\Sigma}_n$ is an approximate solution of the $n$-th iteration, and $\beta$ is a mixing parameter. To improve the convergence, we have used a large $\beta$ when the estimated error 
$|\hat{\Sigma}_{n+1}-\hat{\Sigma}_{n}|$ is linearly converging by iteration 
because it implies that $\hat{\Sigma}_n$ is sufficiently close to the desired solution.

In our numerical calculation, we have used $N_m=64 \times 64$ for the number $N_m$ of $\bm{k}$-mesh and $N_\omega=4096$ for the number $N_\omega$ of the Matsubara frequencies. 
At the temperatures ($T \ge 0.001$) used in calculations, the energy cut-off $(2N_{\omega}+1)\pi T \ge 25.7$ is sufficiently larger than the band width $8t_1=8$. We note that the difference in the free energy between the SC state solution and the normal state solution can be calculated under the fixed $N_{\omega}$\cite{yanase_kinetic_2005}. 
We have also performed a computation using $N_m=128 \times 128$ and/or $N_\omega=8192$, or fixing the energy cut-off instead of $N_\omega$, and have verified that the results are not significantly changed.

In order to investigate the PPB-induced AFM ordering, we have evaluated the transverse AFM susceptibility from the calculated self-energy because the AFM moment is perpendicular to the field in the situation of CeCoIn$_5$ in $\bm{H}\perp \text{c}$\cite{kenzelmann_coupled_2008_kenzelmann_evidence_2010,kumagai_evolution_2011}. The transverse AFM susceptibility 
 is given by 
\begin{equation}
	\label{eq:chit}
	\chi_t(\bm{q})=\frac{\chi_{\pm}^{(0)}(0,\bm{q})}{1-U\chi_{\pm}^{(0)}(0,\bm{q})} .
\end{equation}

Strictly speaking, a magnetic phase transition line can not be calculated in the FLEX approximation 
because the divergence of $\chi_t({\bf q})$ at the transition leads to a numerical instability. Moreover, a true phase transition does not occur in a pure two-dimensional system according to the Mermin-Wagner theorem.
In a quasi two-dimensional system, however, the phase transition can be realized by a weak inter-layer coupling of the layered crystal structure. 
Therefore, the AFM transition line in a quasi two-dimensional system will be determined by the criterion\cite{kino_phase_1998} 
\begin{equation}
	1-U\mathrm{max}_{\bm{q}}[\chi_{\pm}^{(0)}(0,\bm{q})]=J
	\label{eq:AFMtransition}
\end{equation}
where $J$ is the small parameter corresponding to the magnitude of the inter-layer coupling. 
In our calculation, $\chi_{\pm}^{(0)}(0,\bm{q})$ and $\chi_t(\bm{q})$ always take maximum values at $\bm{q}=(\pi,\pi)$ ($=\bm{Q}$) within the resolution of $\bm{k}$-mesh.

\section{
First order superconducting transition
\label{sec:first_order}
}
In this section, we will discuss the first order transition (FOT) occurring between the uniform SC with strong PPB and the normal phases in the high-field regime. This well-known result in the weak-coupling model in the paper of Maki and Tsuneto\cite{maki_pauli_1964}, where the AFM ordering is not incorporated, will be studied here in the strong-coupling approach. For this reason, the two-dimensional case with $J=0$ is assumed here so that the AFM order is suppressed by its fluctuation itself and thus, does not occur at finite temperatures. 
A high-field FOT between the uniform SC and the normal phases is observed in several strong-coupling superconductors where the estimated Maki parameter\cite{maki_effect_1966} 
is large, 
such as CeCoIn$_5$ at intermediate temperature\cite{bianchi_possible_2003} 
and KFe$_2$As$_2$\cite{zocco_pauli-limited_2013}. 

\begin{figure}
\includegraphics[width=83mm]{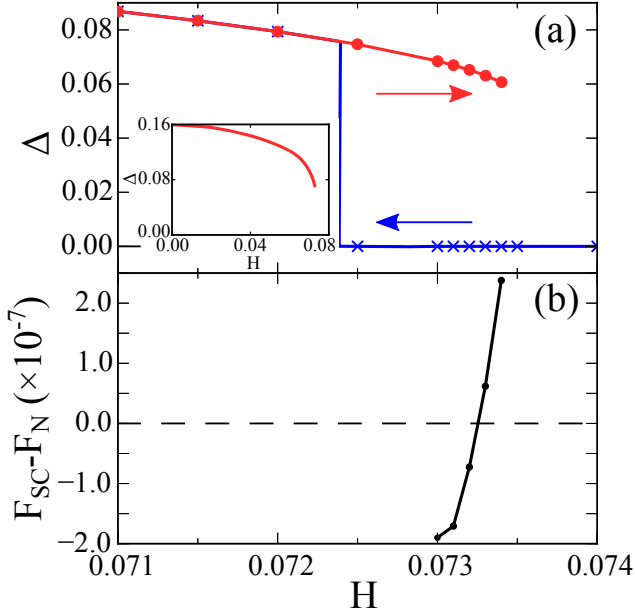}
\caption{
(Color online)
(a)Field dependence of $\Delta$ with elevating (red circle) and lowering (blue cross) the magnetic field $H$. 
Inset shows the field dependence of $\Delta$ in a wider field range.
(b)Field dependence of the free energy difference between the SC state solution and the normal state solution.
The used parameters are $t_2=0.25$, $U=2.8$, $\delta=0.1$, and $T=0.002$
\label{fig:mDk}
}
\end{figure}

A hysteretic behavior of the order parameter is the signature of the FOT. 
Figure \ref{fig:mDk}(a) shows the up and down field sweeps of $\Delta$ at $T=0.002=0.15 T_c$, 
where $\Delta$ is the maximum amplitude of the anomalous self-energy $|\Delta(k) |$ and  can be regarded as the SC order parameter. 
In the down field sweep, $\Delta$ stays zero until $h=0.0724$, which is the $H_{c2}$ obtained by {\it assuming} the second order SC transition;
below this field, the normal state solution ($\Delta=0$) become unstable. 
In the up field sweep, on the other hand, $\Delta$ is still finite in $0.0724<h<0.0734$;
in this field range, both the SC state solution and the normal state solution are the local minima of the free energy functional (eq. (\ref{eq:FE})).
This hysteretic behavior of $\Delta$ indicates that the field-induced SC transition in thermal equilibrium is of first order at this temperature. 

\begin{figure}
\includegraphics[width=83mm]{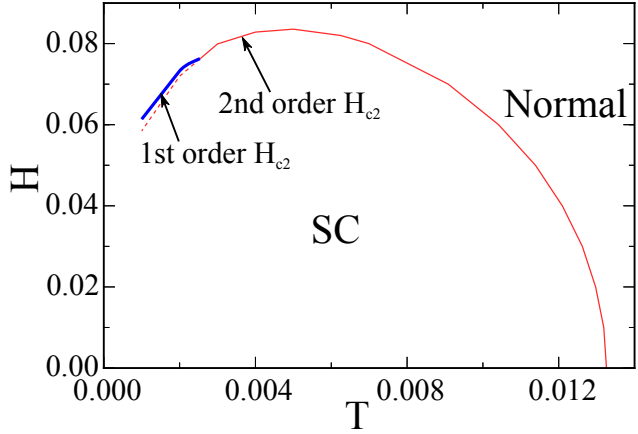}
\caption{
(Color online)
SC transition $H_{c2}(T)$ curve drawn in the $H$-$T$ phase diagram in the two-dimensional case where the AFM never occurs. 
Thin solid (red) curve in $T > 0.0024$ is the $H_{c2}(T)$ curve obtained as a second order transition line, while the bold solid (blue) line in $T < 0.0024$ is the portion of $H_{c2}(T)$ at which the FOT occurs. The portion of the thin dashed (red) curve  in $T < 0.0024$ is not a true transition line. 
The used parameters are $t_2=0.25$, $U=2.8$, $\delta=0.1$, and $J=0$. 
\label{fig:SCphase}
}
\end{figure}

To determine the position of the FOT line, we have compared the free energy of the SC state solution with that of the normal state solution. 
Figure \ref{fig:mDk}(b) shows the difference in the free energy between the SC state and the normal state solutions. It is seen from Fig.\ref{fig:mDk}(b) that the FOT occurs at $h=0.0733$ for $T=0.002$. 
The resulting H-T phase diagram of a superconductor in the Pauli limit and in the two-dimensional case is shown in Fig. \ref{fig:SCphase}. 
At low temperatures ($T < 0.0024$), the $H_{c2}$-transition is of first order, and the portion of the thin dashed (red) line in the same temperature range is not a phase transition line. 
However, the FOT portion of $H_{c2}(T)$ is just slightly larger than the second order portion (thin dashed curve) in the {\it same} temperature range (see the figure). Therefore, in contrast to the corresponding weak-coupling case\cite{maki_pauli_1964}, the first order $H_{c2}(T)$ inevitably \textit{decreases} upon cooling reflecting the downward $T$-dependence of the corresponding second order portion. We note that, in contrast, the first order $H_{c2}(T)$ line \textit{increases} on cooling in any real superconductors showing the FOT on $H_{c2}(T)$ in high fields\cite{bianchi_possible_2003,zocco_pauli-limited_2013}. 

This discrepancy is attributed to the unexpected reduction of $\Delta$ upon increasing the magnetic field $H$ at low temperatures. As seen in the inset of Fig. \ref{fig:mDk}(a), the $\Delta$ value at the transition field is merely the half of that in zero field. Such a large $H$ dependence of $\Delta$ does not appear in the corresponding weak-coupling approach, and may suggest that the PPB effect in the present strong-coupling case is too strong effectively. It is possible that the orbital pair-breaking effect, neglected in the present approach, would weaken the PPB effect and make the downward $H_{c2}(T)$ an upward one. In this sense, the orbital pair-breaking effect seems to play more important roles in the strong-coupling case, and its inclusion has to be left for our future work.

\section{
Antiferromagnetic order induced in $d$-wave superconducting phase 
\label{sec:afm}
}
In this section, we will discuss the AFM ordering induced by the PPB effect in the SC phase in high-field regime. 
For this purpose, we first explain the field-dependence of the calculated transverse AFM susceptibility $\chi_t(\bm{Q})$ (eq. (\ref{eq:chit})) 
in the case with and without SC order. 
Next, in order to find the mechanism of the PPB-induced AFM ordering in the present strong-coupling approach, the normal and anomalous components of the susceptibility are examined in details and compared with the corresponding ones in the weak-coupling model to examine whether or not the mechanism of the AFM ordering in the SC phase is essentially the same as that in the weak-coupling model. 
Furthermore, the H-T phase diagram following from the result of $\chi_t(\bm{Q})$ is presented.

\begin{figure}
\includegraphics[width=83mm]{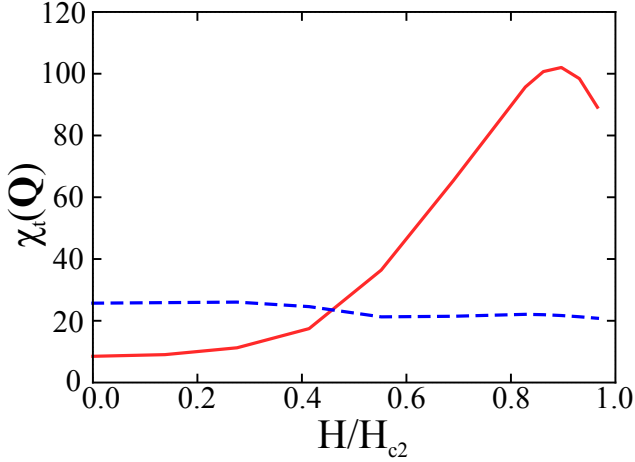}
\caption{
(Color online)
$\chi_t(\bm{Q})$ vs. $H/H_{c2}$ curve in the case with the SC order (solid (red) line) and with no SC order (dashed (blue) line). 
The used parameters are $t_2=0.25$, $U=2.8$, $\delta=0.1$, and $T=0.002$
\label{fig:chit}
}
\end{figure}

Figure \ref{fig:chit} shows the field dependence of $\chi_t(\bm{Q})$ with and without SC order at $T=0.002=0.15 T_c$. 
In low fields, $\chi_t(\bm{Q})$ is suppressed by the SC order, as usually expected in the low-field superconductivity. 
On the other hand, 
$\chi_t(\bm{Q})$ is largely enhanced by the SC order in high fields close to $H_{c2}$, where the PPB effect is strong. 
This result shows that the AFM ordering is enhanced by the PPB effect even if the strong-coupling effects are included.

\begin{figure}
\includegraphics[width=83mm]{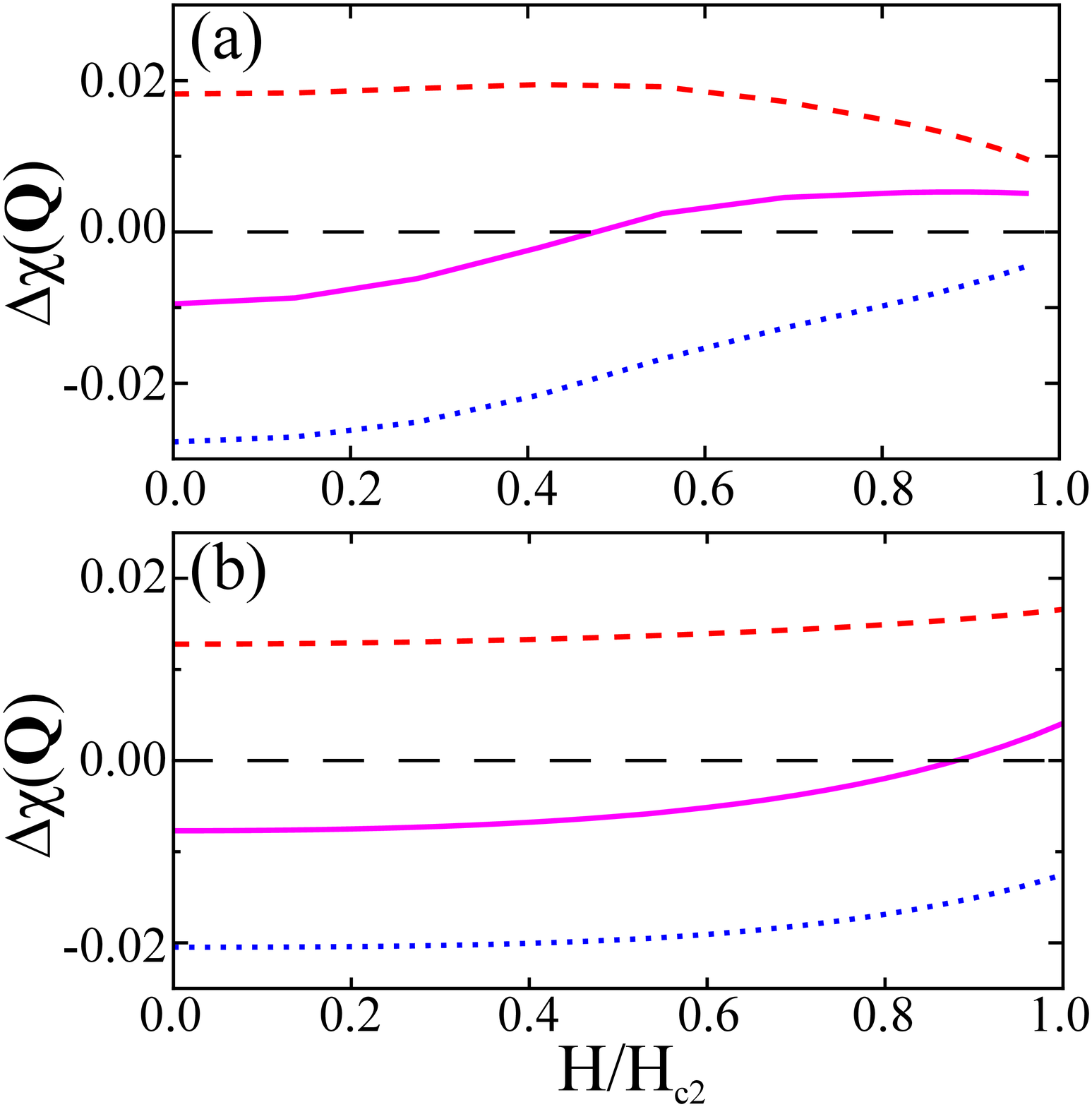}
\caption{
(Color online)
Field dependence of the SC part of each $\chi$ ($\chi_{\pm}^{(0)}(0,\bm{Q})$, $\chi_{\text{GG}}$, or $\chi_{\text{FF}}$) calculated in the strong-coupling model (a) and in the weak-coupling model (b), respectively. 
In both figures, the dotted (blue) line is $\Delta \chi_{\text{GG}}=\chi_{\text{GG}} - \chi_{\text{GG}}(\Delta=0)$, the dashed (red) line is $\Delta \chi_{\text{FF}}=\chi_{\text{FF}}$, and the solid (purple) line is $\Delta \chi_{\pm}^{(0)}(0,\bm{Q})$. 
The used parameters are the same as Fig. \ref{fig:chit}
\label{fig:compare}
}
\end{figure}

Before discussing the mechanism of $\chi_t(\bm{Q})$-enhancement 
in the strong-coupling model, 
we will first summarize here the mechanism of the PPB-induced AFM ordering in the weak-coupling model\cite{hatakeyama_emergent_2011}. 
Since, according to eq.(\ref{eq:chit}), $\chi_{t}(\bm{Q})$ increases with increasing $\chi_{\pm}^{(0)}(\bm{Q})$, 
we will focus on $\chi_{\pm}^{(0)}(\bm{Q})$. 
In the following discussion, $\Delta \chi$ denotes the SC part of $\chi$, that is, the difference between $\chi$ and the corresponding one in the normal 
state, 
where $\chi$ is $\chi_{\pm}^{(0)}(\bm{Q})$, $\chi_{\text{GG}}$, or $\chi_{\text{FF}}$ (defined below). In a SC state, 
$\chi_{\pm}^{(0)}(\bm{Q})$ consists of two contribution, $\chi_{\text{GG}}$ and $\chi_{\text{FF}}$, where 
\begin{equation}
	\chi_{\pm}^{(0)}(0,\bm{Q})=\chi_{\text{GG}}+\chi_{\text{FF}} ,
\end{equation}
\begin{eqnarray}
	\chi_{\text{GG}} &=& -T\sum_k G_{\uparrow}(k)G_{\downarrow}(k+Q), \label{eq:GG}\\
	\chi_{\text{FF}} &=& -T\sum_k F(k)F^{\dagger}(-k-Q) \label{eq:FF} .
\end{eqnarray}
Here, $\chi_{\text{GG}}$ correspond to the contribution from the quasiparticle excitations, while $\chi_{\text{FF}}$ corresponds to the contribution from the Cooper pair condensate. 
The field dependences of $\Delta \chi_{\text{GG}}$ and $\Delta \chi_{\text{FF}}= \chi_{\text{FF}}$ in the $d_{x^2-y^2}$-wave case are plotted in Fig. \ref{fig:compare}(b). 
In the SC state, $\chi_{\text{GG}}$ is suppressed by the presence of the SC energy gap, 
while $\chi_{\text{FF}}$ is enhanced if the SC gap function $\Delta_{\bm{k}}$ changes sign in the $\bm{k}$-space due to the property $\Delta(\bm{k+Q})=-\Delta(\bm{k})$ in the $d_{x^2-y^2}$-wave pairing state, where $\bm{Q} =(\pi,\pi)$. 
Therefore, $\Delta \chi_{\text{GG}}$ is always negative, while $\Delta \chi_{\text{FF}}$ is always positive in Fig. \ref{fig:compare}(b). 
At zero field, $| \Delta \chi_{\text{GG}} |$ is larger than $\Delta \chi_{\text{FF}}$, and thus, as conventionally expected, the AFM ordering is suppressed by SC order 
\cite{kato_superconductivity_1988,konno_superconductivity_1989}. 
In high fields close to $H_{c2}$, on the other hand, $| \Delta \chi_{\text{GG}} |$ decreases 
because the effects of the SC energy gap are reduced by the Zeeman shift of the excitation energy. 
As a result, $|\Delta \chi_{\text{FF}}|$ become larger than $\Delta \chi_{\text{GG}}$, and thus, the AFM ordering is realized by the SC order in the high-field region of the SC phase.

The $\Delta \chi_{\text{GG}}$ and $\Delta \chi_{\text{FF}}$ results in the strong-coupling model are plotted in Fig. \ref{fig:compare}(a), 
and one notices that, by comparing the corresponding curves in the figure (b), $\Delta \chi$ in the strong-coupling case has the following similar feature as that in the weak-coupling model : 
(1) $\Delta \chi_{\text{GG}}$ is always negative, while $\Delta \chi_{\text{FF}}$ is always positive; 
(2) in zero field, $| \Delta \chi_{\text{GG}} |$ is larger than $|\Delta \chi_{\text{FF}}|$, and thus, $\Delta \chi_{\pm}^{(0)}(\bm{Q})$ is negative;
(3) in high fields close to $H_{c2}$, $| \Delta \chi_{\text{GG}} |$ become smaller than $|\Delta \chi_{\text{FF}}|$, and thus, $\Delta \chi_{\pm}^{(0)}(\bm{Q})$ become positive, implying the AFM ordering. 
These features indicate that the mechanism of $\chi_t(\bm{Q})$-enhancement due to the PPB-effect in the strong-coupling model is essentially the same as that in the weak-coupling model. 

In Fig. \ref{fig:chit}, $\chi_t(\bm{Q})$ has a maximum value not at $H_{c2}$ but at the field slightly lower than $H_{c2}$. In contrast, it has a maximum value at $H_{c2}$ where the PPB effect is the strongest in the weak-coupling approach in the corresponding Pauli limit\cite{hatakeyama_emergent_2011}. 
This difference is due to the fact that the SC order parameter is strongly reduced by the PPB effect with increasing the field in the strong-coupling model (see the inset of Fig.1 (a)). 
As already mentioned in relation to the FOT line in Sect. \ref{sec:first_order}, this remarkable reduction of $|\Delta|$ due to the increase of the field may be an artifact of the neglect of the orbital pair-breaking. Therefore, the above-mentioned shift of the maximum of $\chi_t(\bm{Q})$ to a lower field need not to 
be seriously considered. On the other hand, if the orbital pair-breaking is incorporated, the mean field SC transition on $H_{c2}$ tends to become of second order, and consequently, $|\Delta|$ will be reduced on approaching $H_{c2}(T)$ from below. That is, as roles of the orbital pair-breaking become important 
enough, inevitably the field at which the AFM ordering is the strongest is shifted to a lower field than $H_{c2}(0)$.  
A similar feature has been obtained in the weak-coupling model in the case 
with the second order $H_{c2}$\cite{hatakeyama_emergent_2011}, 
where the SC order parameter is reduced with increasing the field by the orbital pair-breaking effect. 
We note that this feature cannot be explained by the picture argued in the paper of Kato et al.\cite{kato_antiferromagnetic_2011} that the nesting between the PPB-induced quasiparticle pockets leads to the AFM ordering, 
because a reduction of the SC gap amplitude leads to a larger quasiparticle pockets and thus, a better nesting. 
Therefore, it is necessary to include the contribution from the Cooper pair condensate in order to describe the AFM ordering occurring close to $H_{c2}(0)$. 

\begin{figure}
\includegraphics[width=83mm]{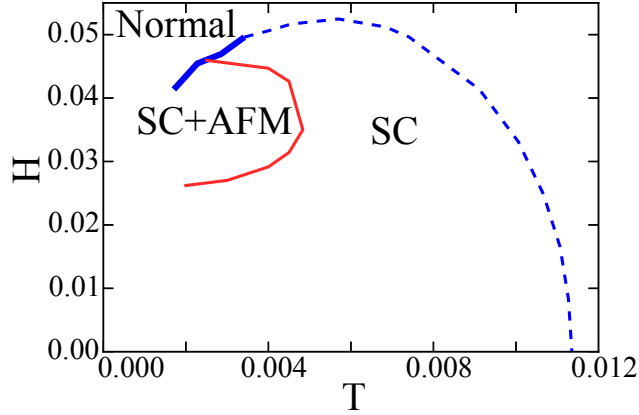}
\caption{
(Color online)
Example of possible H-T phase diagrams. The solid (red) line is 
the AFM transition line, the dashed (blue) line is the second order $H_{c2}$ line, and the thick solid (blue) line is the possible first order $H_{c2}$ line on which the hysteretic behavior of $\Delta$ is observed. 
The used parameters are $t_2=0.28$, $U=2.8$, $\delta=0.1$ and $J=0.045$.
\label{fig:AFMphase}
}
\end{figure}

Figure \ref{fig:AFMphase} shows a possible H-T phase diagram in the strong-coupling model.
Here, the AFM transition line is determined by the criterion given in eq. (\ref{eq:AFMtransition}).
In this phase diagram, the AFM order is present only in the high-field region of the SC phase and not present in the high-field normal state.
Furthermore, the AFM order disappears together with the first order $H_{c2}$-transition at low temperatures. 
These features are the same as those obtained in the weak-coupling model in the Pauli limit\cite{hatakeyama_emergent_2011} and are comparable with those seen in CeCoIn$_5$. 
The spatial modulation of the gap amplitude $|\Delta|$ peculiar to the FFLO state, which is believed to be realized only in the HFLT phase of CeCoIn$_5$\cite{kumagai_evolution_2011}, is not included in Fig. \ref{fig:AFMphase}. 
On the other 
hand, the weak-coupling results have shown that the AFM ordered region is significantly expanded by the FFLO modulation of $|\Delta|$\cite{hatakeyama_emergent_2011}. 
In this manner, the AFM order present only in the HFLT phase of CeCoIn$_5$ can be understood based on the mechanism of the PPB-induced AFM ordering.

\section{
Strong-coupling effects on antiferromagnetic ordering
\label{sec:detail}
}

In this section, we will investigate an influence of the strong-coupling effects on the PPB-induced AFM ordering in detail. 

First, we will review the conventional point of view on the strong-coupling effects on the electronic properties. 
The anomalous self-energy $\Delta(k)$ includes strong-coupling effects on the SC order. This 
$\Delta(k)$ is different from the SC order parameter in the weak-coupling model in the following senses:
(1) $\Delta(k)$ is dependent on the Fermion Matsubara frequency; (2) $|\Delta(k)|$ is enhanced as a consequence of the strong pairing interaction. 
On the other hand, the normal self-energy $\Sigma_{\sigma}(k)$ includes the strong-coupling effect on the quasiparticle. 
As shown later, $\Delta\chi_{\pm}^{(0)}(\bm{Q})$ is affected by only the low-energy excitation.
Because the low-energy excitation can be described by the Fermi liquid picture,
the resulting strong-coupling effects on the quasiparticle may be represented by the following parameters: 
(1) the quasiparticle damping rate $\gamma$, and (2) the renormalization factor $a$, which is related to the effective mass $m^*$ as $m^{*}=m/a$. Below, 
we will investigate effects of these parameters 
on $\Delta\chi_{\pm}^{(0)}(\bm{Q})$. 

\begin{figure}
\includegraphics[width=83mm]{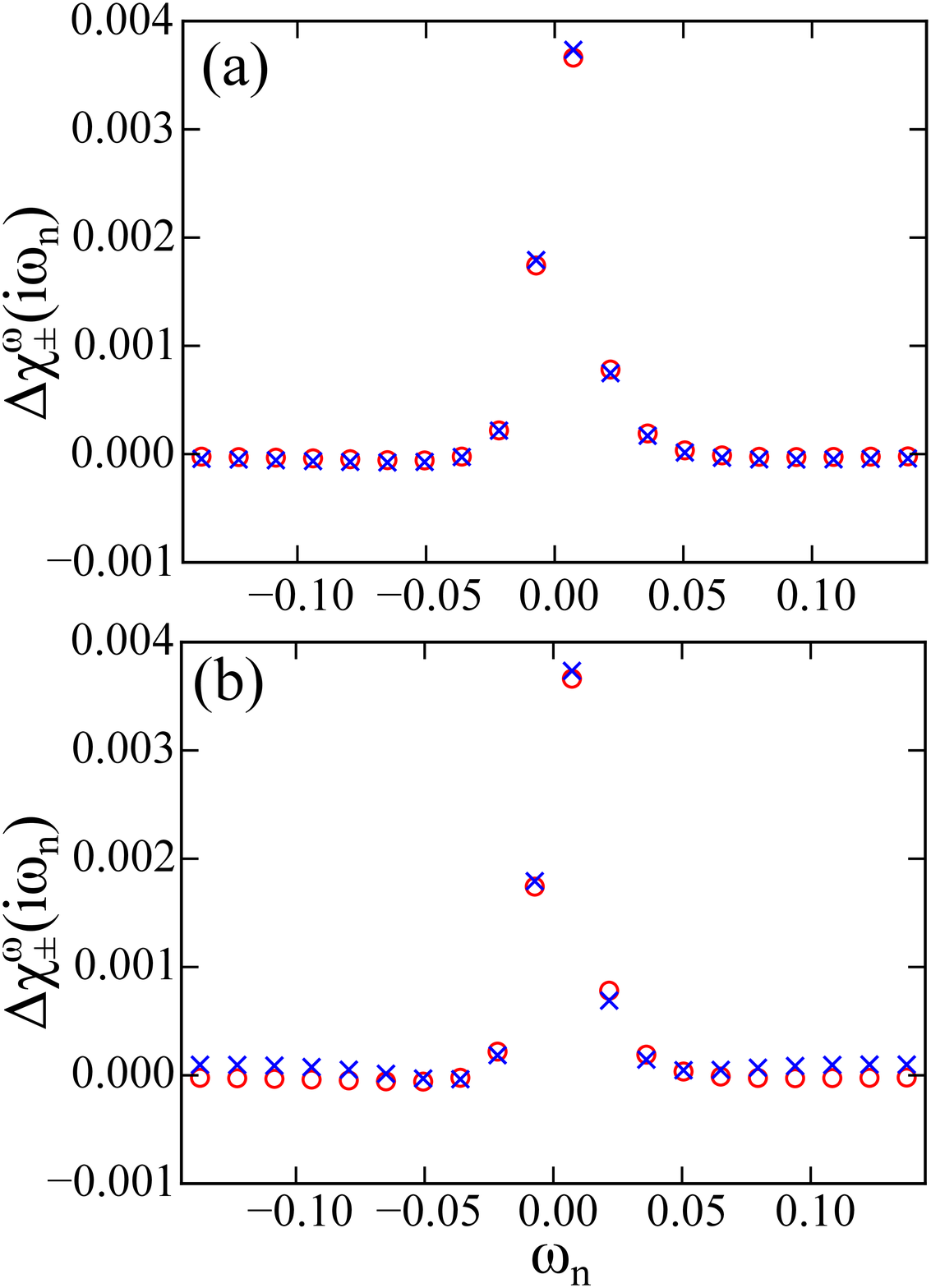}
\caption{
(Color online)
Frequency dependence of $\Delta \chi_{\pm}^{\omega}$. 
In both figure, red circles express $\Delta \chi_{\pm}^{\omega}$ with the full frequency dependence of the self-energy. 
In (a), blue cross symbols express $\Delta \chi_{\pm}^{\omega}$ obtained by assuming that  $\Delta(k)$ is frequency-independent. 
In (b), blue crosses express $\Delta \chi_{\pm}^{\omega}$ obtained using the Fermi liquid approximation for $\Sigma_{\sigma}(k)$ (see the text). 
The used parameters are $t_2=0.28$, $U=2.8$, $\delta=0.1$, $T=0.0023$ and $H=0.035$. 
\label{fig:diff}
}
\end{figure}

In order to examine the effect of the frequency-dependence of the self-energy on $\Delta\chi_{\pm}^{(0)}(\bm{Q})$, $\chi_{\pm}^{(0)}(\bm{Q})$ will be expressed as the sum over the Matsubara frequencies 
\begin{equation}
	\chi_{\pm}^{(0)}(\bm{Q})=\sum_{\omega_n} \chi_{\pm}^{\omega}(i\omega_n) ,
\end{equation}
where
\begin{equation}
	\chi_{\pm}^{\omega}(i\omega_n)=-T\sum_{\bm{k}} \left[ G_{\uparrow}(i\omega_n,\bm{k})G_{\downarrow}(i\omega_n,\bm{k+Q})+F(i\omega_n,\bm{k})F^{\dagger}(-i\omega_n,\bm{-k-Q}) \right] .
\end{equation}
Note that $\chi_{\pm}^{\omega}(i\omega_n)$ depends only on $\hat{\Sigma}(i\omega_n,\bm{k})$ at the same Matsubara frequency $\omega_n$ through 
$G_{\sigma}(i\omega_n,\bm{k})$, $F(i\omega_n,\bm{k})$, and $F^{\dagger}(i\omega_n,\bm{k})$. 

To examine the effect of the frequency dependence of $\Delta(k)$ on $\Delta \chi_{\pm}^{(0)}(\bm{Q})=\chi_{\pm}^{(0)}({\bf{Q}}) - \chi_{\pm}^{(0)}({\bf{Q}})|_{\Delta=0}$, 
we also calculate $\Delta \chi_{\pm}^{\omega}$ by 
assuming that $\Delta(k)$ is frequency-independent, i.e., substituting $\Delta'(\bm{k})=\Delta(i\pi T,\bm{k})$ for $\Delta(k)$.
It is compared to $\Delta \chi_{\pm}^{\omega}$ with full frequency dependence of $\Delta(k)$ in Fig. \ref{fig:diff}(a); differences between 
these two plots are negligibly small, indicating that 
the frequency dependence of $\Delta(k)$ does not play major roles on $\Delta \chi_{\pm}^{(0)}(\bm{Q})$. In other words, it is the amplitude of $\Delta(k)$ which mainly determines the effects of superconductivity on $\Delta \chi_{\pm}^{(0)}(\bm{Q})$. 

We also examined the effect of the normal self-energy on $\Delta \chi_{\pm}^{(0)}(\bm{Q})$ by calculating $\Delta \chi_{\pm}^{\omega}$.
In Fig. \ref{fig:diff}, $\Delta \chi_{\pm}^{\omega}$ (red circle) decays in the energy scale $| \omega_n | \sim 0.04$, 
which corresponds to the SC excitation gap $E_g \sim a |\Delta(i\pi T,\bm{k}=(0,\pi))| \sim 0.04$.
This suggests that only the low-energy excitations $\omega<E_g \ll t_1$ are relevant to $\Delta\chi_{\pm}^{(0)}(\bm{Q})$.
Therefore, it is expected that the low-frequency part of the normal self-energy can be approximated in terms of its Fermi liquid form to calculate $\Delta\chi_{\pm}^{(0)}(\bm{Q})$; 
\begin{equation}
	\Sigma_{\sigma}(i\omega_n,\bm{k}) \sim c_{\sigma}(\bm{k})+(1-a_{\sigma}^{-1}(\bm{k}))i\omega_n, 
	\label{eq:FLapprox}
\end{equation}
where $c_{\sigma}(\bm{k})\in \mathbb{R}$ represents the correction to the dispersion relation by the interaction (it can be absorbed to the definition of $\epsilon_{\bm{k}}$),
and $a_{\sigma}(\bm{k})$ is the renormalization factor.
Here, the imaginary part of $c_{\sigma}(\bm{k})$, which represents the quasiparticle damping $\gamma$, is ignored because the quasiparticle damping is 
 negligible at low temperature and low energy in the Fermi liquid theory.
We calculate $\Delta \chi_{\pm}^{\omega}$ in the Fermi liquid approximation and compare it to 
$\Delta \chi_{\pm}^{\omega}$ with the full frequency dependence of $\Sigma_{\sigma}(i\omega_n)$ in Fig. \ref{fig:diff}(b).
Since these two plots are quantitatively close to each other, it is found that the effects of the normal self-energy on $\Delta \chi_{\pm}^{\omega}$ are well approximated by the Fermi liquid approximation.
That is, the effect of $\Sigma_{\sigma}(k)$ on  $\Delta \chi_{\pm}^{(0)}(\bm{Q})$ appears mainly through the renormalization factor.

The above-mentioned results show that the strong-coupling effect on the $\Delta\chi_{\pm}^{(0)}(\bm{Q})$ is mainly dependent on 
the amplitude of the SC order parameter and the quasiparticle renormalization.
Because it is difficult to vary these parameters independently in the strong-coupling model, 
we examined the influence of these parameters on $\Delta\chi_{\pm}^{(0)}(\bm{Q})$ by performing the following model calculation {\it based on the weak-coupling model}.
In this model, the Fermi liquid approximation for $\Sigma_{\sigma}(k)$ is employed and the frequency dependence of $\Delta(k)$ is ignored;
\begin{eqnarray}
	\Sigma_{\sigma}(k) &\sim & (1-a^{-1})i\omega_n \label{eq:model_sigma} \\
	\Delta(k) &\sim & \Delta w_{\bm{k}} \label{eq:model_D} ,
\end{eqnarray}
where the renormalization factor $a$ is assumed to be a positive constant, $\Delta$ is the SC order parameter which is independent of $\omega_n$, and $w_{\bm{k}}=\cos(k_x)-\cos(k_y)$ is the $d_{x^2-y^2}$-wave pairing function. 
Here, $c({\bm{k}})$ has been absorbed to the definition of 
$\epsilon_{\bm{k}}$. 
The SC order parameter $\Delta$ and the first order transition curve $H_{c2}(T)$ is determined from the following free energy in the weak-coupling BCS model 
with the substitution $\omega_n\rightarrow \omega_n/a$ (by eq. (\ref{eq:model_sigma})) : 
\begin{equation}
	F(\Delta)=\frac{\Delta^2}{\lambda}-T\sum_{\omega_n>0}\sum_{\bm{k}}\ln \left[ 
	\frac{((\omega_n/a)^2+[\epsilon_{\bm{k}}]^2+|\Delta w_{\bm{k}}|^2-h^2)+4\omega_n^2h^2/a^2}
	{((\omega_n/a)^2+[\epsilon_{\bm{k}}]^2-h^2)+4\omega_n^2h^2/a^2} \right] ,
\end{equation}
where $\lambda$ denote the strength of the pairing interaction in the $d_{x^2-y^2}$-wave channel. 

\begin{figure}[tb]
\includegraphics[width=83mm]{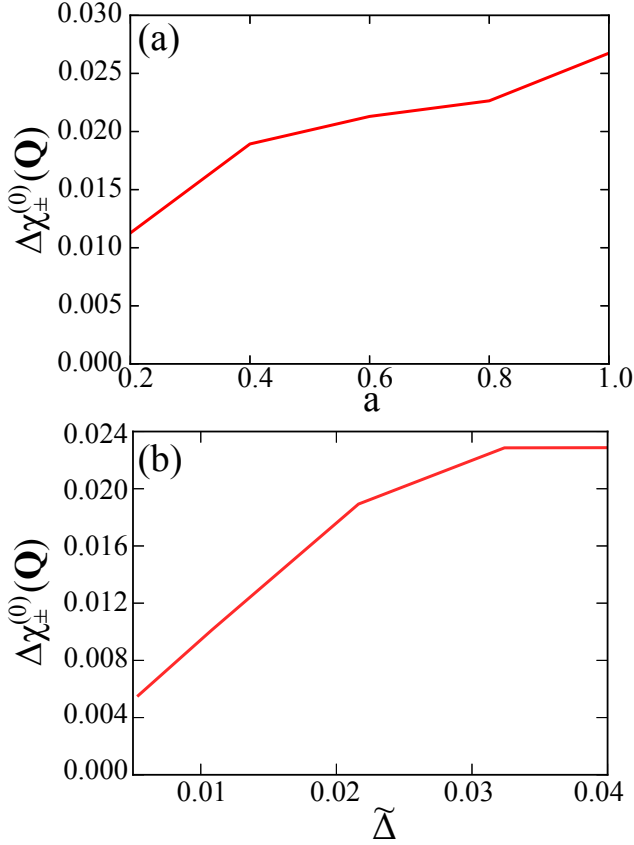}
\caption{
(Color online)
$\Delta \chi_{\pm}^{(0)}(\bm{Q})$ at the FOT line $H_{c2}$ vs. (a) the renormalization factor $a$ and (b) the SC excitation gap $\tilde{\Delta}$ 
in the model calculation based on the weak-coupling model. 
The used parameter are $t_2=0.28$, $\delta=0.1$, and $T/T_c=0.1$. 
The fixed value $\tilde{\Delta}=0.02$ and $a=0.4$ are used in (a) and (b), respectively. 
\label{fig:zD}
}
\end{figure}

The result of the model calculation is shown in Fig. \ref{fig:zD}.
Figure \ref{fig:zD} (a) plots the $a$ dependence of $\Delta \chi_{\pm}^{(0)}(\bm{Q})$ at $H_{c2}$ with a fixed SC excitation gap $\tilde{\Delta} \equiv a\Delta$ ($=0.02$). 
This figure shows that $\Delta \chi_{\pm}^{(0)}(\bm{Q})$ decreases as $a$ decreases, i.e., as the quasiparticle renormalization becomes stronger. 
Figure \ref{fig:zD} (b) plots the $\tilde{\Delta}$ dependence of $\Delta \chi_{\pm}^{(0)}(\bm{Q})$ at $H_{c2}$ with a fixed $a$ ($=0.4$). 
This figure shows that $\Delta \chi_{\pm}^{(0)}(\bm{Q})$ increases as $\tilde{\Delta}$ increases, i.e., as the pairing interaction becomes stronger. 
Because a stronger AFM fluctuations leads to a larger renormalization and a stronger pairing interaction, 
these two effects play competitive roles with each other on $\Delta \chi_{\pm}^{(0)}(\bm{Q})$ when the AFM fluctuation is enhanced. 

As shown above, 
as the AFM fluctuation becomes strong, the AFM ordering is suppressed by the strong quasiparticle renormalization, while it is enhanced by the 
large SC order parameter. However, the model calculation cannot determine which of these two effects plays more dominant roles when the AFM fluctuation becomes strong. 
Therefore, we will discuss in the next section the dependences of the PPB-induced AFM ordering on the microscopic parameters in the strong-coupling model.

\section{
Electronic parameter dependences of PPB-induced AFM ordering
\label{sec:parameter}
}

In this section, we will discuss dependences of the AFM ordering occuring near $H_{c2}(0)$ on the microscopic details. In the present Hubbard model (eq. (\ref{eq:Hamiltonian})), we have three microscopic parameters, the repulsive interaction strength $U$, the next-nearest-neighbor hoping integral $t_2$, and the filling fraction $n$ ($=1-\delta$). It is expected that, increasing $U$ enhances the AFM fluctuation without changing the dispersion relation, while that increasing $t_2$ makes the Fermi surface nesting worse and thus, suppresses the AFM fluctuation. In addition, increasing $\delta$ by hole doping makes the nesting worse by expanding the hole Fermi surface in $\bm{k}$-space and thus, suppresses the AFM fluctuation. 

\begin{figure}
\includegraphics[width=83mm]{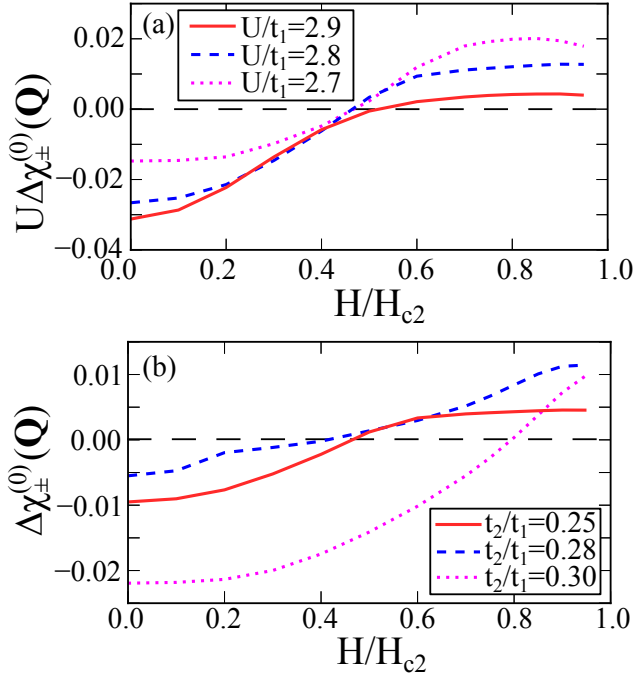}
\caption{
(Color online)
(a) $U$ dependence of $U\Delta\chi^{(0)}_{\pm}(\bm{Q})$ vs. $H/H_{c2}$ line, and (b) $t_2$ dependence of $\Delta\chi^{(0)}_{\pm}(\bm{Q})$ vs. $H/H_{c2}$ line.
The used parameters are $t_2=0.25$, $\delta=0.1$ and $T/T_c=0.15$ in (a), and $U=2.8$, $\delta=0.1$ and $T/T_c=0.15$ in (b).
\label{fig:chit_irr}
}
\end{figure}

Figure \ref{fig:chit_irr}(a) shows $U$ dependence of $U\Delta \chi_{\pm}^{0}(\bm{Q};h)$ at $T/T_c=0.15$.  
In this figure, $U\Delta \chi_{\pm}^{0}(\bm{Q})$ close to $H_{c2}(0)$ decreases when the AFM fluctuation is enhanced (i.e. $U$ is increased). 
Due to the result of the model calculation in Sect. \ref{sec:detail}, 
it is concluded that the influence on $\Delta \chi_{\pm}^{0}(\bm{Q})$ of the quasiparticle renormalization is larger than that of the amplitude of the SC order parameter, 
and thus, $\Delta \chi_{\pm}^{0}(\bm{Q})$ is suppressed when the AFM fluctuation is enhanced.

\begin{figure}
\includegraphics[width=83mm]{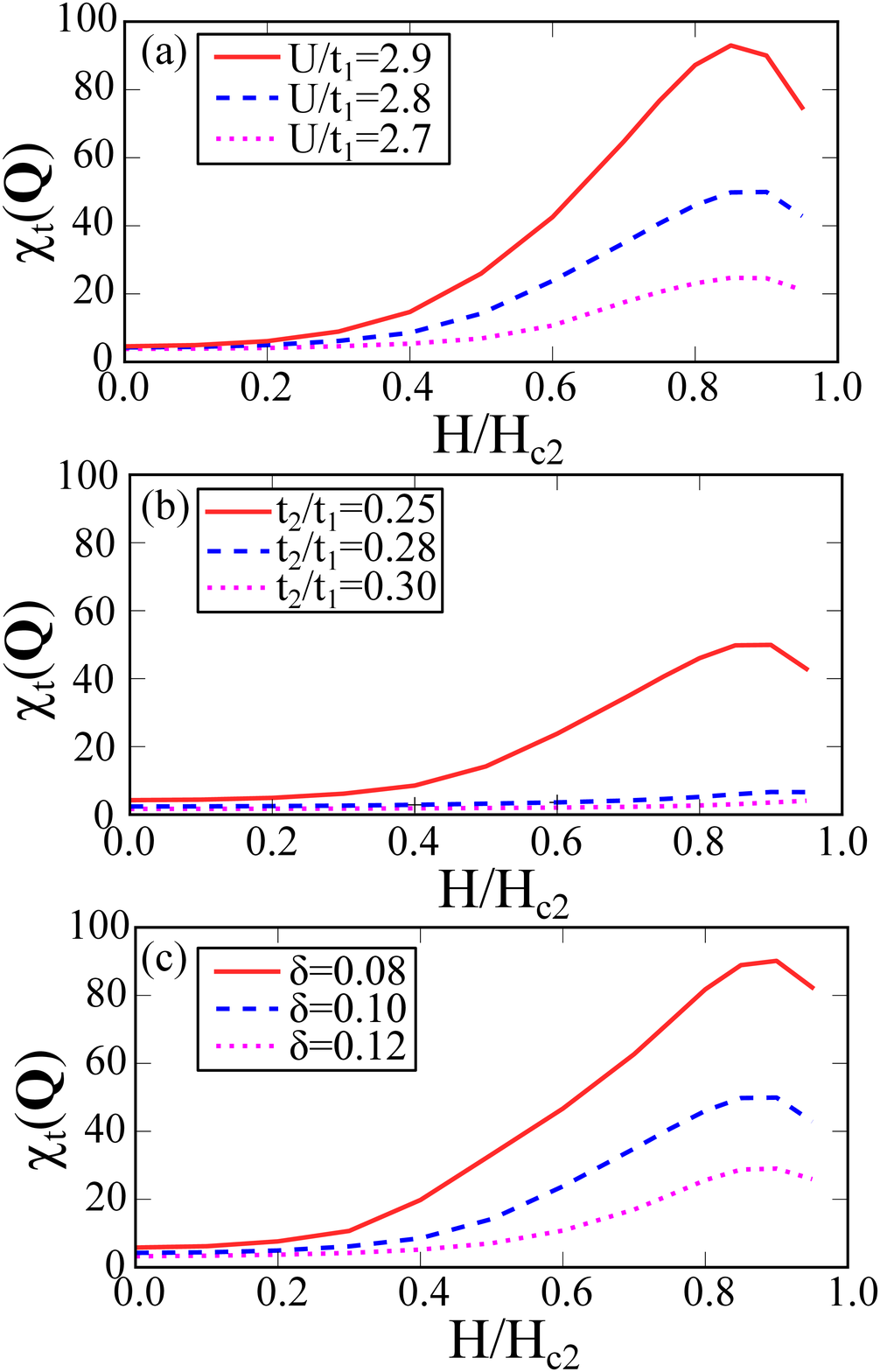}
\caption{
(Color online)
(a) $U$, (b) $t_2$, and (c) $\delta$ dependence of $\chi_t(\bm{Q})$ vs. $H/H_{c2}$ line.
The used parameters are 
$t_2=0.25$, $\delta=0.1$ and $T/T_c=0.15$ in (a),
$U=2.8$, $\delta=0.1$ and $T/T_c=0.15$ in (b),
and $U=2.8$, $t_2=0.25$ and $T/T_c=0.15$ in (c).
\label{fig:chi_all}
}
\end{figure}

Next, we will discuss the parameter dependences of $\chi_t(\bm{Q})$ in order to understand the effect of proximity to the AFM critical point {\it in the 
normal state}. In Fig. \ref{fig:chi_all}, 
$U$, $t_2$, and $\delta$ dependences of $\chi_t(\bm{Q};h)$ are plotted. 
Contrary to the above result, 
the peak of $\chi_t(\bm{Q};h)$ is largely enhanced when the AFM fluctuation is enhanced 
(i.e. when $U$, $\delta$ is increased, or $t_2$ is decreased). 
This enhancement of  $\chi_t(\bm{Q})$ is due to the increase of the Stoner enhancement factor 
\begin{equation}
\alpha = \frac{1}{1-U\chi_{\pm}^{0}(\bm{Q})}.
\end{equation}
The difference between $\chi_t(\bm{Q})$ and the corresponding one in the normal state, $\Delta\chi_t(\bm{Q})$, is expressed as
\begin{equation}
	\Delta\chi_t(\bm{Q}) \sim \alpha^2 U\chi_{\pm}^{0}(\bm{Q}) \Delta\chi_{\pm}^{0}(\bm{Q})
\end{equation}
when $\alpha \gg 1$ and $U\Delta\chi_{\pm}^{(0)}(\bm{Q}) \ll 1$, where $\chi_{\pm}^{0}(\bm{Q})$ can be regarded as the expression in the normal state. 
Since $\chi_t(\bm{Q})$ {\it in the normal state} is given by $\chi_t(\bm{Q})=\alpha\chi_{\pm}^{0}(\bm{Q})$, 
$\Delta\chi_t(\bm{Q})$, which is proportional to $\alpha^2$, is more strongly enhanced than its normal value ($\propto \alpha$) on approaching the AFM-QCP. Even if taking account of the parameter dependences of $\Delta \chi_{\pm}^{0}(\bm{Q})$ in addition to this enhancement of $\chi_t(\bm{Q})$ {\it in the SC state}, enhancement of the AFM ordering due to the Stoner factor dominates 
over suppression of the AFM ordering due to the quasiparticle renormalization close to the AFM-QCP. 

Next, we will discuss the parameter dependence of the field region with enhanced AFM fluctuation, in which $\chi_t(\bm{Q})$ is enhanced by the presence of the SC state (i.e. $\Delta\chi_{\pm}^{0}(\bm{Q})>0$). 
The AFM order appearing only inside a SC phase can be realized within this field region. 
In Fig. \ref{fig:chit_irr}(b), 
the field region with enhanced AFM fluctuation is expanded with increasing the strength of the AFM fluctuation. In this figure, it is seen by comparing the case with strong AFM fluctuation (e.g., the case with relatively smaller $t_2$-values such as $2.8$ and $2.5$) 
to the case with $t_2=3.0$ leading to a weak AFM fluctuation. 
This expansion of the AFM region follows from the reduction of the SC order parameter $\Delta$ due to the strong PPB effect. 
In the case with the weak AFM fluctuation, the field dependence of $\Delta$ is found to be quite weak at low temperatures as in the weak-coupling model.
In the case with the strong AFM fluctuation, on the other hand, $\Delta$ at low temperatures is found to be reduced largely by the field, as discussed in Sect. \ref{sec:first_order}. 
Because the strength of the PPB effect is determined by the ratio between the Zeemen energy and the SC gap amplitude in the Pauli limit where $\alpha_M\rightarrow \infty$ (see Sect. \ref{sec:introduction}), 
the PPB effect becomes relatively strong even in low fields due to the reduction of $\Delta$. 
Therefore, the AFM field region is extended to lower fields as the AFM fluctuation becomes stronger. 

Contrary to this, the opposite tendency to Fig. \ref{fig:chit_irr}(b) is seen in Fig. \ref{fig:chit_irr}(a), where 
the AFM field region is reduced by an increase of $U$ which also leads to the increase of $\chi_t(\bm{Q})$. This feature in Fig. \ref{fig:chit_irr}(a) may be attributed to the reduction of the quasiparticle renormalization factor $a$. 
In the model calculation in Sect. \ref{sec:detail}, $\chi^{(0)}_{\pm}$ with an isotropic dispersion relation is scaled by the renormalization factor $a$ as,
\begin{equation}
	\chi^{(0)}_{\pm}(T,h;a,\Delta,\delta_{\text{IC}})=a\chi^{(0)}_{\pm}(T,ah;1,a\Delta,a\delta_{\text{IC}}),
	\label{eq:a_scale}
\end{equation}
where $\delta_{\text{IC}}$ denotes the imperfect nesting of the Fermi surface, which is defined by $\xi_{\bm{k+Q}}=-\xi_{\bm{k}}+\delta_{\text{IC}}$.
Based on eq. (\ref{eq:a_scale}), it is found that 
the effective imperfect nesting $\tilde{\delta}_{\text{IC}}=a\delta_{\text{IC}}$ decreases as $a$ is reduced by the strong AFM fluctuations. 
Since, in the PPB-induced AFM ordering, a more perfect nesting leads to a shrinkage of the AFM field region\cite{hatakeyama_emergent_2011}, it implies a shrinkage of the AFM field region due to an enhancement of the AFM fluctuation. 

\section{
Summary and discussion
\label{sec:summary}
}
In the present paper, we have discussed the strong-coupling effect on the mechanism of the PPB-induced AFM ordering, shown in the previous works focusing on the HFLT phase of CeCoIn$_5$\cite{ikeda_antiferromagnetic_2010}, on the basis of the theoretical analysis using the FLEX approximation for the two-dimensional Hubbard model with the Zeeman energy. 
In Sect. \ref{sec:afm}, it has been found that the PPB-induced AFM ordering is realized in the strong-coupling model, and that its mechanism 
is essentially the same as that in the weak-coupling model. It has been argued that the PPB-induced AFM ordering occurs even if the orbital depairing effect is included\cite{hatakeyama_emergent_2011} and can be realized over a wide range of systems 
which have the strong PPB effect and a sign change of $\Delta_{\bm{k}}$ in $\bm{k}$-space (e.g., $d$-wave superconductors). 
In Sect.\ref{sec:afm}, the enhancement of the AFM ordering below $H_{c2}$ has been explained as a result of the reduction of $\Delta$ close to $H_{c2}$ (Fig. \ref{fig:chit}). 
This result shows that the contribution from the Cooper pair condensate to $\chi_{\pm}^{(0)}(\bm{Q})$ is important for the PPB-induced AFM ordering. 
The mechanism of the PPB-induced AFM ordering might also be applied to the iron-based superconductors 
with strong PPB effect and $s_{\pm}$-wave pairing\cite{zocco_pauli-limited_2013}, if the gap function $\Delta_{\bm{k}}$ changes its sign with a translation in $\bm{k}$-space such as $\bm{k} \to \bm{k}+\bm{Q}$ with $\bm{Q}=(\pi,0)$. 
Moreover, The AFM-QCP located near $H_{c2}(0)$ observed in many $d$-wave superconductors can be explained 
based on the enhancement of the AFM fluctuations due to the PPB effect close to $H_{c2}$. 

In Sect. \ref{sec:detail} and \ref{sec:parameter},
we have investigated influences of the strong-coupling effects on the PPB-induced AFM ordering when the AFM-QCP is approached. 
When the AFM fluctuation is enhanced, the SC part of the {\it irreducible} susceptibility $\Delta \chi_{\pm}^{(0)}(\bm{Q})$ is suppressed by the quasiparticle renormalization (Fig. \ref{fig:zD}(a)), while it is enhanced by the large SC order parameter induced by the strong pairing interaction (Fig. \ref{fig:zD}(b)). It is found that, close to the AFM-QCP, the effect of the quasiparticle renormalization on $\Delta \chi_{\pm}^{(0)}(\bm{Q})$ is larger than that of the large SC order parameter (Fig. \ref{fig:chit_irr}). However, it is important to note that, in the observable AFM susceptibility $\chi_t(\bm{Q})$, the Stoner enhancement factor depending on various parameters is also included. In fact, it has been shown that the effect of the Stoner enhancement is dominant over that of the quasiparticle renormalization.
As a result, the PPB-induced AFM ordering is enhanced when the AFM-QCP is approached by applying pressure or doping.
In addition, it has been shown that 
the AFM ordered region can become wider as the AFM fluctuation is enhanced (Fig. \ref{fig:chit_irr}(b)). 
Therefore, the field-induced AFM order present only inside a SC phase is expected to be realized close to the AFM-QCP. 
In fact, in pressured CeRhIn$_5$, the AFM transition line in the H-T phase diagram is contracted toward a field slightly below $H_{c2}(0)$ 
as the pressure is increased toward $p_c \sim 2.5\text{GPa}$\cite{park_hidden_2006} corresponding to the location of the AFM-QCP. 
The situation similar to CeRhIn$_5$ might be realized in many of heavy electron and high-$T_c$ superconductors with a strong PPB effect 
because the SC phase in those materials is often realized close to the AFM-QCP, and the coexistence of AFM and SC order is observed in some of 
these materials. 

In Sect. \ref{sec:first_order}, we have showed that the first order transition between the uniform SC and the normal states 
is realized at low temperatures and in high fields 
in the strong-coupling model, although, unexpectedly, the resulting $H_{c2}(T)$ decreases upon cooling at the lowest temperatures. To improve this disagreement on $H_{c2}(T)$ curve with the experimental data\cite{bianchi_possible_2003}, inclusion of the orbital pair-breaking seems to be necessary. Together with this, a consistent inclusion of the FFLO ordering in the present calculation will be left for a future work.

\begin{acknowledgment}
One of authors (Y.H.) thanks H. Ikeda and Y. Yanase for the valuable discussions. This work was supported by Grants-in-Aid for Scientific Research (No.25400368) from MEXT, Japan. Y.H. is supported by JSPS Research Fellowship for Young Scientists.
\end{acknowledgment}

\appendix
\section{
calculation of the second order $H_{c2}(T)$
}
In this appendix, we will demonstrate the detail of our calculation of the second order $H_{c2}(T)$ between the uniform SC and the normal states. 
The normal self-energy is calculated based on the self-consistent equation obtained from eq. (\ref{eq:selfeq_S}) by substituting $\Delta(k)=0$. On the other hand, the linearized Eliashberg equation is obtained by linearizing eq. (\ref{eq:selfeq_D}) with respect to $\Delta(k)$ and becomes 
\begin{eqnarray}
	\label{eq:linearized}
	\lambda\Delta(k) &=& -T\sum_{q} \left[ \left( \frac{U}{1-U^2\chi_{\uparrow}^{(0)}(q)\chi_{\downarrow}^{(0)}(q)} \right) F(k+q)
		\right. \nonumber\\ 
		&& \left. +\frac{U^2\chi_{\pm}^{(0)}(q)}{1-U\chi_{\pm}^{(0)}(q)}F(-k-q) \right] ,
\end{eqnarray}
where $\lambda$ denote the eigenvalue. 
Here, $F(k)$ is also linearized with respect to $\Delta(k)$ and becomes 
\begin{equation}
	\label{eq:linearF}
	F(k)=G_{\uparrow}(k)G_{\downarrow}(-k)\Delta(k) .
\end{equation}
The eigenvalue equation (\ref{eq:linearized}) is solved to obtain the largest eigenvalue $\lambda_{\text{max}}$,
and then the second order $H_{c2}(T)$ is determined by finding the point where $\lambda_{\text{max}}=1$.


\begin{thebibliography}{10}

\bibitem{bianchi_possible_2003}
A.~Bianchi, R.~Movshovich, C.~Capan, P.~G. Pagliuso, and J.~L. Sarrao: Phys.
  Rev. Lett. {\bfseries 91} (2003) 187004.

\bibitem{fulde_superconductivity_1964_larkin_inhomogeneous_1965}
P.~Fulde and R.~A. Ferrell: Phys. Rev. {\bfseries 135} (1964) A550;
A.~I. Larkin and Y.~N. Ovchinnikov: Sov. Phys. {JETP} {\bfseries 20} (1965)
  762.

\bibitem{tokiwa_anisotropic_2008}
Y.~Tokiwa, R.~Movshovich, F.~Ronning, E.~D. Bauer, P.~Papin, A.~D. Bianchi,
  J.~F. Rauscher, S.~M. Kauzlarich, and Z.~Fisk: Phys. Rev. Lett. {\bfseries
  101} (2008) 037001.

\bibitem{tokiwa_anomalous_2010}
Y.~Tokiwa, R.~Movshovich, F.~Ronning, E.~D. Bauer, A.~D. Bianchi, Z.~Fisk, and
  J.~D. Thompson: Phys. Rev. B {\bfseries 82} (2010) 220502.

\bibitem{kumagai_evolution_2011}
K.~Kumagai, H.~Shishido, T.~Shibauchi, and Y.~Matsuda: Phys. Rev. Lett.
  {\bfseries 106} (2011) 137004.

\bibitem{kenzelmann_coupled_2008_kenzelmann_evidence_2010}
M.~Kenzelmann, T.~Str{\"a}ssle, C.~Niedermayer, M.~Sigrist, B.~Padmanabhan,
  M.~Zolliker, A.~D. Bianchi, R.~Movshovich, E.~D. Bauer, J.~L. Sarrao, and
  J.~D. Thompson: Science {\bfseries 321} (2008) 1652;
M.~Kenzelmann, S.~Gerber, N.~Egetenmeyer, J.~L. Gavilano, T.~Str{\"a}ssle,
  A.~D. Bianchi, E.~Ressouche, R.~Movshovich, E.~D. Bauer, J.~L. Sarrao, and
  J.~D. Thompson: Phys. Rev. Lett. {\bfseries 104} (2010) 127001.

\bibitem{park_hidden_2006}
T.~Park, F.~Ronning, H.~Q. Yuan, M.~B. Salamon, R.~Movshovich, J.~L. Sarrao,
  and J.~D. Thompson: Nature {\bfseries 440} (2006) 65.

\bibitem{bianchi_avoided_2003}
A.~Bianchi, R.~Movshovich, I.~Vekhter, P.~Pagliuso, and J.~Sarrao: Phys. Rev.
  Lett. {\bfseries 91} (2003) 257001.

\bibitem{paglione_field-induced_2003}
J.~Paglione, M.~Tanatar, D.~Hawthorn, E.~Boaknin, R.~Hill, F.~Ronning,
  M.~Sutherland, L.~Taillefer, C.~Petrovic, and P.~Canfield: Phys. Rev. Lett.
  {\bfseries 91} (2003) 246405.

\bibitem{singh_probing_2007}
S.~Singh, C.~Capan, M.~Nicklas, M.~Rams, A.~Gladun, H.~Lee, J.~F. {DiTusa},
  Z.~Fisk, F.~Steglich, and S.~Wirth: Phys. Rev. Lett. {\bfseries 98} (2007)
  057001.

\bibitem{zaum_towards_2011}
S.~Zaum, K.~Grube, R.~Sch{\"a}fer, E.~D. Bauer, J.~D. Thompson, and
  H.~v.~L{\"o}hneysen: Phys. Rev. Lett. {\bfseries 106} (2011) 087003.

\bibitem{park_field-induced_2010}
T.~Park, Y.~Tokiwa, F.~Ronning, H.~Lee, E.~D. Bauer, R.~Movshovich, and J.~D.
  Thompson: Phys. Status Solidi B {\bfseries 247} (2010) 553.

\bibitem{dong_field-induced_2011}
J.~K. Dong, H.~Zhang, X.~Qiu, B.~Y. Pan, Y.~F. Dai, T.~Y. Guan, S.~Y. Zhou,
  D.~Gnida, D.~Kaczorowski, and S.~Y. Li: Phys. Rev. X {\bfseries 1} (2011)
  011010.

\bibitem{tokiwa_quantum_2011}
Y.~Tokiwa, P.~Gegenwart, D.~Gnida, and D.~Kaczorowski: Phys. Rev. B {\bfseries
  84} (2011) 140507.

\bibitem{honda_effect_2008}
F.~Honda, R.~Settai, D.~Aoki, Y.~Haga, T.~D. Matsuda, N.~Tateiwa, S.~Ikeda,
  Y.~Homma, H.~Sakai, Y.~Shiokawa, E.~Yamamoto, A.~Nakamura, and Y.~Onuki: J.
  Phys. Soc. Jpn. {\bfseries 77} (2008) 339.

\bibitem{hatakeyama_emergent_2011}
Y.~Hatakeyama and R.~Ikeda: Phys. Rev. B {\bfseries 83} (2011) 224518.

\bibitem{ikeda_antiferromagnetic_2010}
R.~Ikeda, Y.~Hatakeyama, and K.~Aoyama: Phys. Rev. B {\bfseries 82} (2010)
  060510.

\bibitem{maki_effect_1966}
K.~Maki: Phys. Rev. {\bfseries 148} (1966) 362.

\bibitem{petrovic_heavy-fermion_2001}
C.~Petrovic, P.~G. Pagliuso, M.~F. Hundley, R.~Movshovich, J.~L. Sarrao, J.~D.
  Thompson, Z.~Fisk, and P.~Monthoux: J. Phys.: Condens. Matter {\bfseries 13}
  (2001) L337.

\bibitem{aoki_unconventional_2007}
D.~Aoki, Y.~Haga, T.~D. Matsuda, N.~Tateiwa, S.~Ikeda, Y.~Homma, H.~Sakai,
  Y.~Shiokawa, E.~Yamamoto, A.~Nakamura, R.~Settai, and Y.~{\=O}nuki: J. Phys.
  Soc. Jpn. {\bfseries 76} (2007) 063701.

\bibitem{moriya_spin_2000}
T.~Moriya and K.~Ueda: Adv. Phys. {\bfseries 49} (2000) 555.

\bibitem{monthoux_self-consistent_1994}
P.~Monthoux and D.~J. Scalapino: Phys. Rev. Lett. {\bfseries 72} (1994) 1874.

\bibitem{sakurazawa_magnetic-field-induced_2005}
K.~Sakurazawa, H.~Kontani, and T.~Saso: J. Phys. Soc. Jpn. {\bfseries 74}
  (2005) 271.

\bibitem{baym_self-consistent_1962}
G.~Baym: Phys. Rev. {\bfseries 127} (1962) 1391.

\bibitem{adachi_effects_2003}
H.~Adachi and R.~Ikeda: Phys. Rev. B {\bfseries 68} (2003) 184510.

\bibitem{ikeda_impurity-induced_2010}
R.~Ikeda: Phys. Rev. B {\bfseries 81} (2010) 060510(R).

\bibitem{maki_pauli_1964}
K.~Maki and T.~Tsuneto: Prog. Theor. Phys. {\bfseries 31} (1964) 945.

\bibitem{yanase_kinetic_2005}
Y.~Yanase and M.~Ogata: J. Phys. Soc. Jpn. {\bfseries 74} (2005) 1534.

\bibitem{kino_phase_1998}
H.~Kino and H.~Kontani: J. Phys. Soc. Jpn. {\bfseries 67} (1998) 3691.

\bibitem{zocco_pauli-limited_2013}
D.~A. Zocco, K.~Grube, F.~Eilers, T.~Wolf, and H.~v. L{\"o}hneysen: Phys. Rev.
  Lett. {\bfseries 111} (2013) 057007.

\bibitem{kato_superconductivity_1988}
M.~Kato and K.~Machida: Phys. Rev. B {\bfseries 37} (1988) 1510.

\bibitem{konno_superconductivity_1989}
R.~Konno and K.~Ueda: Phys. Rev. B {\bfseries 40} (1989) 4329.

\bibitem{kato_antiferromagnetic_2011}
Y.~Kato, C.~D. Batista, and I.~Vekhter: Phys. Rev. Lett. {\bfseries 107} (2011)
  096401.

\bibitem{capan_fermi_2010}
C.~Capan, Y.-J. Jo, L.~Balicas, R.~G. Goodrich, J.~F. {DiTusa}, I.~Vekhter,
  T.~P. Murphy, A.~D. Bianchi, L.~D. Pham, J.~Y. Cho, J.~Y. Chan, D.~P. Young,
  and Z.~Fisk: Phys. Rev. B {\bfseries 82} (2010) 035112.

\bibitem{knebel_quantum_2008}
G.~Knebel, D.~Aoki, J.-P. Brison, and J.~Flouquet: J. Phys. Soc. Jpn.
  {\bfseries 77} (2008) 114704.

\end{thebibliography}

\end{document}